\documentclass{article}

\usepackage[final,nonatbib]{nips_2018}

\usepackage[utf8]{inputenc} % allow utf-8 input
\usepackage[T1]{fontenc}    % use 8-bit T1 fonts
\usepackage{hyperref}       % hyperlinks
\usepackage{url}            % simple URL typesetting
\usepackage{booktabs}       % professional-quality tables
\usepackage{amsfonts}       % blackboard math symbols
\usepackage{nicefrac}       % compact symbols for 1/2, etc.
\usepackage{microtype}      % microtypography
\usepackage{comment}
\usepackage{graphicx}
\usepackage{color}
\usepackage{amsmath}
\usepackage{subcaption}
\usepackage{booktabs}
\usepackage{multirow}
\usepackage{listings}
\usepackage{graphicx,wrapfig,lipsum}

\usepackage[skip=0.5\baselineskip]{caption}

\graphicspath{ {./images/} }

\usepackage{ifthen}
\usepackage[normalem]{ulem} % for \sout
\usepackage{xcolor}
\usepackage{amssymb}

\newboolean{showedits}
\setboolean{showedits}{false} % toggle to show or hide edits
\ifthenelse{\boolean{showedits}}
{
	\newcommand{\del}[1]{\textcolor{red}{\sout{#1}}} % please delete
}{
	\newcommand{\del}[1]{} % please delete
	
}

\newboolean{showcomments}
\setboolean{showcomments}{false} % ATTN: THIS TURNS ON THE ANNOTATIONS
\setboolean{showcomments}{true} % ATTN: THIS TURNS ON THE ANNOTATIONS
\newcommand{\id}[1]{$-$Id: scgPaper.tex 32478 2010-04-29 09:11:32Z oscar $-$}

\ifthenelse{\boolean{showcomments}}
{\newcommand{\nbc}[3]{
 {\colorbox{#3}{\bfseries\sffamily\scriptsize\textcolor{white}{#1}}}
 {\textcolor{#3}{\sf\small$\blacktriangleright$\textit{#2}$\blacktriangleleft$}}}
 }
{\newcommand{\nbc}[3]{}
 \renewcommand{\del}[1]{} % please delete
 }

\definecolor{ibcolor}{rgb}{0.9,0.5,0}
\definecolor{cfcolor}{rgb}{0,0.5,0.9}
\definecolor{tdcolor}{rgb}{1.0,0,0}

\title{Iroko: A Framework to Prototype Reinforcement Learning for Data Center 
Traffic Control}

\author{
    Fabian Ruffy\thanks{Equal contribution}~~\thanks{University of British Columbia} \\
        \texttt{fruffy@cs.ubc.ca} \\
    \And
    Michael Przystupa\footnotemark[1]~~\footnotemark[2] \\
    \texttt{bot267@cs.ubc.ca} \\
    \And
    Ivan Beschastnikh\footnotemark[2] \\
    \texttt{bestchai@cs.ubc.ca} \\
    }

\begin{document}

\maketitle

\begin{abstract}

Recent networking research has identified that data-driven congestion control 
(CC) can be more efficient than traditional CC in TCP. Deep reinforcement 
learning (RL), in particular, has the potential to learn optimal network 
policies.
However, RL suffers from instability and over-fitting, deficiencies which so far
render it unacceptable for use in datacenter networks.
In this paper, we analyze the requirements for RL to succeed in the datacenter 
context. We present a new emulator, Iroko, which we developed to support
different network topologies, congestion control algorithms, and deployment 
scenarios. Iroko interfaces with the OpenAI gym toolkit, which allows for fast 
and fair evaluation of different RL and traditional CC algorithms under the 
same conditions. We present initial benchmarks on three deep RL algorithms 
compared to TCP New Vegas and DCTCP. Our results show that these algorithms 
are able to learn a CC policy which exceeds the performance of TCP New Vegas on 
a dumbbell and fat-tree topology. We make our emulator open-source and publicly 
available:
{\small\url{https://github.com/dcgym/iroko}}.

\end{abstract}

%%%%%%%%%%%%%%%%%%%%%%%%%%%%%%%%%%%%%%%%%%%%%%%%%%%%%%
\section{Introduction}
\label{sec:intro}
Reinforcement learning (RL) has seen a surge of interest in the networking community.
Recent contributions include data-driven flow control for wide-area
networks~\cite{pcc}, job scheduling~\cite{ml_resource}, and cellular
congestion control~\cite{ml_cellular}.  A particularly promising
domain is the data center (DC) as many DC networking challenges can be formulated as
RL problems~\cite{deepconfig}. Researchers have used RL to
address a range of DC tasks such as routing~\cite{learning_route,
  ml_routing}, power management~\cite{ml_power}, and traffic
optimization~\cite{auto}.

Adhering to the objective of maximizing future rewards~\cite{rl}, RL has the 
potential to learn anticipatory policies. Data center CC, can benefit from this 
feature, as current DC flow control 
protocols and central schedulers are based on the fundamentally reactive TCP 
algorithm~\cite{d3,fastpass}. While many techniques are designed to respond to 
micro-bursts or flow collisions as quickly as possible,  they are not capable 
of preemptively identifying and avoiding these events~\cite{perc, expresspass}. 
Any time flows collide, packets and goodput is lost. Given the 
availability of data and the range of RL algorithms, CC is an excellent 
match for RL.

However, a lack of generalizability~\cite{Marc2017AUnified,
  Raghu2017CanDeep,marivate, Whiteson2011ProtectingAE,
  Chiyuan2018AStudyOn} and reproducibility~\cite{Henderson2018DeepRL}
makes RL an unacceptable choice for DC operators, who expect
stable, scalable, and predictable behavior.  Despite these
limitations, RL is progressing quickly in fields such as autonomous
driving~\cite{Sallab2017DeepRL} and
robotics~\cite{Jens2013RLRoboticsSurvey}. These domains exhibit
properties similar to DC control problems: both deal with a large
input space and require continuous output actions.  Decisions have to
be made rapidly (on the order of microseconds) without compromising
safety and reliability.

What these fields have, and what current DC research is missing, is a
common platform to compare and evaluate techniques.  RL benchmark
toolkits such as the OpenAI gym~\cite{Brockman2016Open} or RLgarage
(formerly RLlab)~\cite{Duan2016Benchmarking} foster innovation and
enforce a common standards framework.  In the networking space, the
Pantheon project~\cite{pantheon} represents a step in this
direction. It provides a system to compare CC solutions for wide area
networks. No such framework currently exists for DCs, partially
because topology and traffic patterns are often considered private and
proprietary~\cite{traffic}.

We contribute \textbf{Iroko}, a DC emulator to understand the requirements and 
limitations of applying RL in DC
networks. Iroko interfaces with the OpenAI
gym~\cite{Brockman2016Open} and offers a way to fairly evaluate
centralized and decentralized RL algorithms against conventional
traffic control solutions.

As preliminary evaluation, we compare three existing RL algorithms
in a dumbbell and a fat-tree topology. We also discuss the design of our 
emulator, as well as limitations and challenges of using reinforcement learning 
in the DC context.

%\begin{figure}
%	\centering
%	\includegraphics[width=1\linewidth]{parking_lotv2}
%	\caption{}
%	\label{fig:parkinglot}
%\end{figure}

%%%%%%%%%%%%%%%%%%%%%%%%%%%%%%%%%%%%%%%%%%%%%%%%%%%%%%
\section{The Data Center as RL Environment}
\label{sec:dc_env}
A major benefit of datacenters is the flexibility of deployment choices. Since a
DC operator has control over all host and hardware elements, they can 
manage traffic at fine granularity.
An automated agent has many deployment options in the data center. Common 
techniques to mitigate congestion include admission 
control~\cite{fastpass,expresspass}, load-balancing of network 
traffic~\cite{hedera,microte}, queue management~\cite{queue_jump,aqm}, or 
explicit hardware modification~\cite{dctcp,conga}. As TCP is inherently a 
self-regulating, rate-limiting protocol, our emulator uses admission control 
to moderate excess traffic. 

\subsection{Patterns of Traffic}
In order for an algorithm to operate proactively, it needs to be able to 
predict future network state.
It is unclear how trainable DC traffic truly is, as public data is few and far 
between. However, prior work indicates that repeating patterns do
exist~\cite{msr_dc,fb_dc,ml_throughput,microte,learning_tcp}.

Although the use of online-learning algorithms for the Internet is 
controversial~\cite{throwdown}, PCC~\cite{pcc} and Remy~\cite{remy} have 
demonstrated that congestion control algorithms that evolve from trained data 
can compete with or even exceed conventional, manually tuned algorithms.
The idea of drawing from \textit{past} data to learn for the \textit{future} is 
attractive. It is particularly viable in data centers, which are fully 
observable, exhibit specific application patterns~\cite{msr_dc}, and operate on 
recurrent tasks.

If DC traffic is sufficiently predictable, it is possible to design a 
proactive algorithm which forecasts the traffic matrix in future iterations, 
and accordingly adjusts host sending rates.
We agree with prior debate that a local, greedily optimizing 
algorithm may not be capable of achieving this goal~\cite{throwdown}.
Instead, utility needs to be maximized by leveraging global knowledge, either 
over a centralized solution such as FastPass~\cite{fastpass} or distributed in 
terms of a message passing solution such as PERC~\cite{perc}.

\subsection{Decentralized vs Centralized Control}

Control techniques are either centralized or decentralized.
TCP, for example, is a decentralized control algorithm. Each host 
optimizes traffic based on its local control policy and tries to maximize its 
own utility. Recent data-driven variants of TCP include PCC~\cite{pcc} or 
Copa~\cite{copa}, which actively learn an optimal traffic 
policy based on local metrics. 

A decentralized control scheme offers the advantage of scalability and reliability when handling long flows with extend round trip times (RTTs). DCs in contrast typically exhibit short and bursty flows with low RTTs, which limits convergence ~\cite{perc}. This limitation frequently leads to inefficient flow utilization or instantaneous 
queue build up~\cite{fastpass,expresspass, d3}.

In contrast to the decentralized control of TCP, a centralized policy can use 
the global information to efficiently manage network nodes. Recent examples 
include the FastPass~\cite{fastpass} arbiter, the Auto~\cite{auto} traffic 
manager, or the Hedera~\cite{hedera} flow 
scheduler.
A major concern of centralized systems is signaling latency delay and 
limitations in processing power.
However, a central scheme has the potential to plan ahead and asynchronously 
grants hosts traffic guarantees based on its current anticipated model of the 
network. MicroTE~\cite{microte}, which optimizes routing by predicting the next 
short-term traffic matrix, represents an instantiation of this model.

\subsection{Sources of Information}
The available options for network data acquisition in a DC range from switch 
statistics, application flows, job deployment monitoring, or even explicit 
application notifications.

Our algorithms use metrics from the transport layer and below, 
which have traditionally been used in TCP congestion control algorithms. 
Modeling an objective function based only on congestion signals is a tried and 
tested approach. Remy~\cite{remy} and PCC~\cite{pcc} have demonstrated that it 
is possible to dynamically learn and improve the congestion function from 
simple network feedback.

Theoretically, it is possible to query for switch buffer occupancy, packet 
drops, port utilization, active flows, and RTT. End-hosts can 
provide metrics in goodput, latency, jitter, and individual loss. 

As improvement over the packet- and delay-based TCP, relative increases in RTT 
have been effectively used as a signal in congestion avoidance 
research~\cite{timely, bbr}. Queue length in switch interfaces is a discrete 
value and precedes an increase in RTT, making it an equivalent congestion 
metric to RTT increase.

Measured throughput represents the current utilization of the network and acts 
as a metric of the actual utility for an active policy (a network without 
traffic has zero queuing, after all). A one-hot encoding of active TCP/UDP 
flows per switch-port can serve as basis to identify network patterns.

%%%%%%%%%%%%%%%%%%%%%%%%%%%%%%%%%%%%%%%%%%%%%%%%%%%%%%
\section{Emulator Design}
%%%%%%%%%%%%%%%%%%%%%%%%%%%%%%%%%%%%%%%%%%%%%%%%%%%%%%

%%%%%%%%%%%%%%%%%%%%%%%%%%%%%%%%%%%%%%%%%%%%%%%%%%%%%%
\begin{figure}
  \centering
  \includegraphics[width=\linewidth]{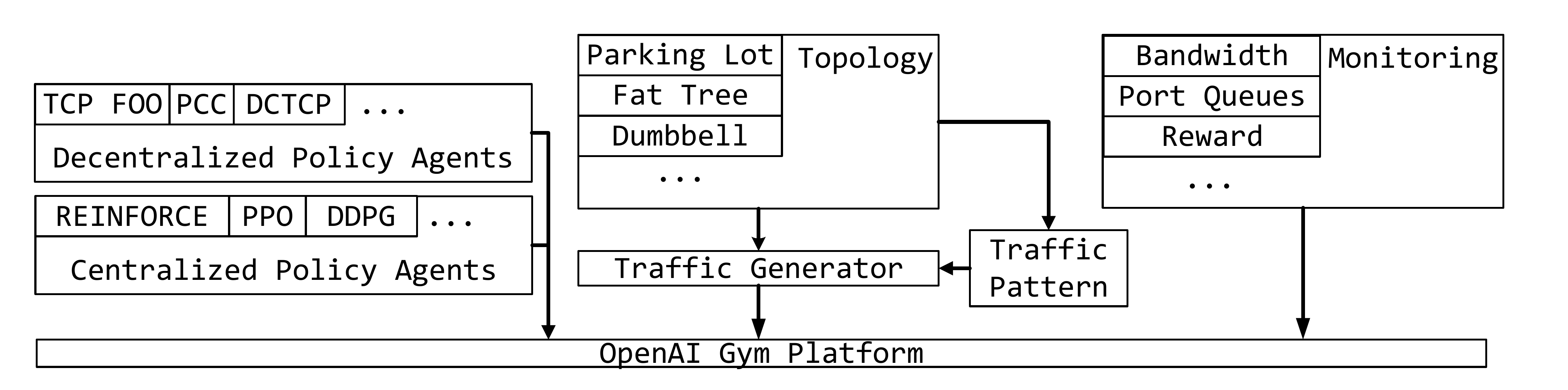}
  \caption{Architecture of the Iroko emulator.}
  \label{fig:iroko_design}
\vspace{-10pt}
\end{figure}

%%%%%%%%%%%%%%%%%%%%%%%%%%%%%%%%%%%%%%%%%%%%%%%%%%%%%%

We designed Iroko to be extensible and modular. All DC
information is abstracted away from the RL agent, providing
flexibility in data acquisition and modeling.  The testing environment
is assembled by combining a set of core components.  This includes a
network topology, a traffic generator, monitors for collecting data
center information, and an agent to enforce the congestion policy (see
Figure~\ref{fig:iroko_design}). The emulator's flexibility allows it
to support centralized arbiters as well as decentralized, host-level
CC approaches. In general, decentralized agents represent traditional TCP 
algorithms such as DCTCP~\cite{dctcp}, TIMELY~\cite{timely}, or PCC~\cite{pcc}, 
while centralized agents operate as RL policies.The emulator also supports 
hybrid deployments, which could operate as multi-agent systems as described 
in~\cite{deepconfig}. The monitors feed information to the agent or record data 
for evaluation. The topology defines the underlying infrastructure and traffic 
patterns that the DC hosts will send.

Two major components of our platform are the Mininet~\cite{mininet} real-time 
network emulator and the Ray~\cite{Lian2018RLLib} project. We use Mininet to 
deploy a virtual network topology and Ray to integrate RL algorithms with the 
emulator. While Mininet is entirely virtual and is limited in its ability to 
generalize DC traffic, it is capable of approximating real traffic 
behavior. Mininet has been effectively used as a platform for larger 
emulation frameworks~\cite{Csoma2014ESCAPE,medicine}.

\subsection{Defining the Environment State}

We have opted for a centralized DC management strategy. All RL agents operate on 
the global view of the network state. 

Iroko deploys monitors that collect statistics from switches in the network and 
store them as a $d \times n$ traffic matrix. This matrix models the data center 
as a list of $n$ ports with $d$ network features.
The agent only uses switch buffer occupancy, interface utilization, and active 
flows as the environment state\footnote{We presume that switches in a real 
DC can reliably provide these statistics.}.
This matrix can be updated on the scale of milliseconds, which is sufficient to 
sample the majority of DC flows~\cite{fb_dc, auto}.

\subsection{The Agent Actions}

Our control scheme specifies the percentage of the maximum allowed
bandwidth each host can send. We represent this action set as a vector
$\Vec{a} \in \mathbb{R}^{n}$ of dimensions equal to the number of host
interfaces.

Each dimension $a_{i}$ represents the percentage\footnote{It is the user's 
responsibility to squash these values to the appropriate range of 0 to 1.0.} of 
maximum bandwidth allocated to the corresponding host by the following 
operation:
\begin{equation}
\text{bw}_{i} \leftarrow bw_{max} * a_{i} \quad  \forall \quad  \text{i} \in \text{hosts}
\end{equation}

Similar to FastPass~\cite{fastpass}, the granularity of this allocation scheme 
can be extended to a per-flow allocation, with a minimum bandwidth 
guarantee per host. For now, we leave it to the agents to estimate the 
best percentage allocation according to the reward. In an ideal instantiation 
of a DC under this system, packet-loss will only rarely, if ever, occur, and 
will minimally impact the network utilization.

\subsection{Congestion Feedback Reward Function}

Choosing an appropriate reward function is crucial for the agent to
learn an optimal policy. Inappropriately defined reward can lead to 
unexpected behavior~\cite{faulty_reward}. The Iroko emulator allows the 
definition of arbitrary reward functions based on the provided input state. In 
our initial setup, we have decided to minimize switch 
bufferbloat~\cite{bufferbloat}. The goal is to reduce the occurrence of queuing 
on switch interfaces, as it indicate congestion and inefficient flow 
distribution.

We follow a common trade-off model which is inspired by recent work 
on TCP CC optimization~\cite{learning_tcp}:
\begin{equation}
R \leftarrow \sum_{i \in hosts} \underbrace{bw_{i} / bw_{max}}_{\rm \text{bandwidth reward}} - 
\underbrace{\text{ifaces}}_{weight} \cdot \underbrace{(q_{i} / q_{max})^2}_{\rm 
	\text{queue penalty}}  - \underbrace{\text{std}}_{dev penalty}
\end{equation}
This equation encourages the agent to find an optimal, fair bandwidth allocation for each host while minimizing switch queues. In the equation, \textit{bandwidth
  reward} is the current network utilization and is the only positive
reward, while \textit{queue penalty} is the current queue size of
switch ports weighted by the number of interfaces.
The \textit{dev penalty} penalizes for actions with high standard deviation to 
ensure allocation fairness and to mitigate host starvation.

\begin{figure}[h]
	%\hspace*{-1.4cm}
	\label{fig:results}
	\begin{subfigure}[h]{\textwidth}
		\includegraphics[width=\textwidth]{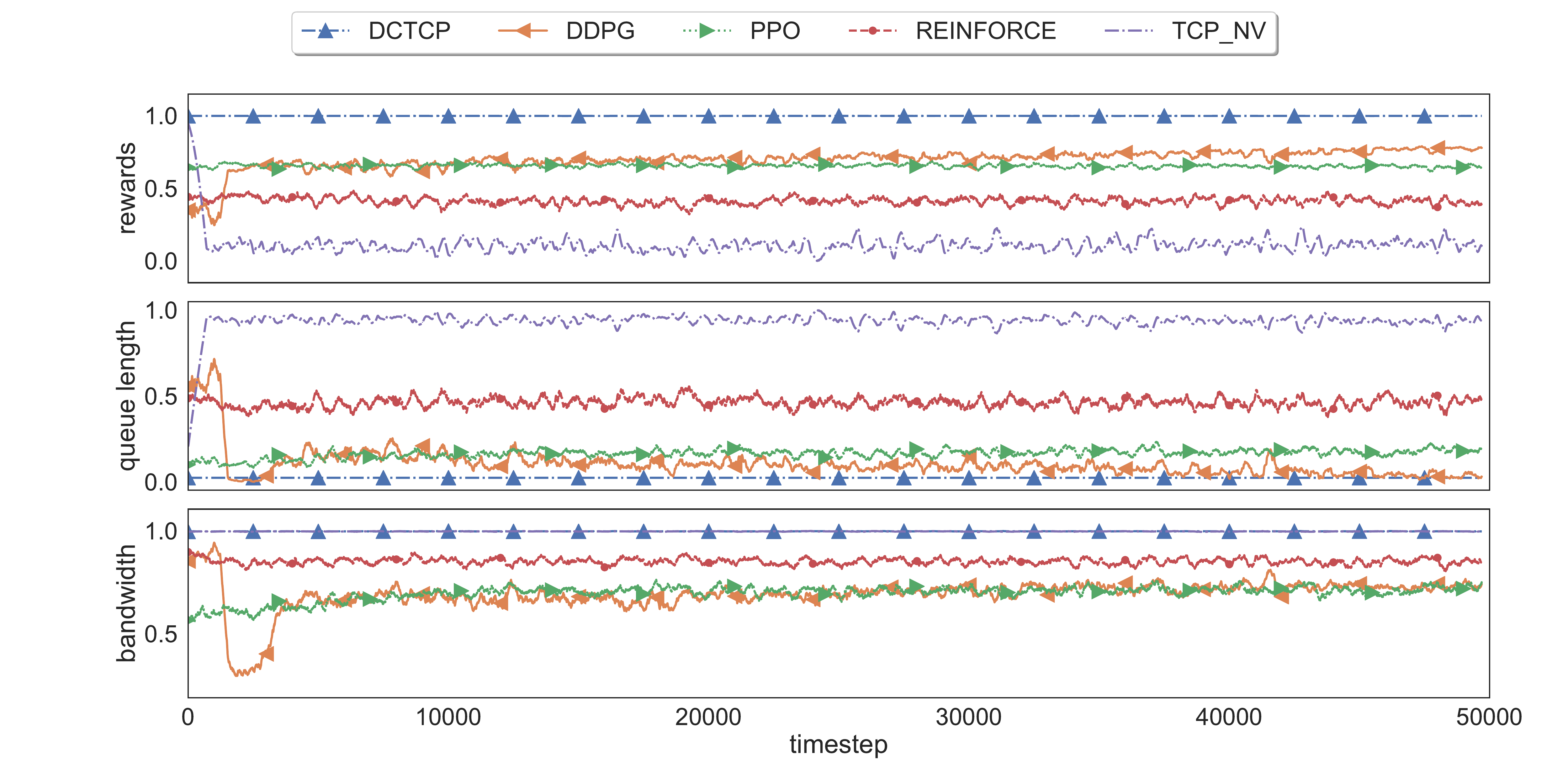}
		\caption{Algorithm performance on a UDP dumbbell topology.}
		\label{fig:dumbbell_udp}
	\end{subfigure}
	
	%\centering
	%
	\begin{subfigure}[h]{\textwidth}
		\includegraphics[width=\textwidth]{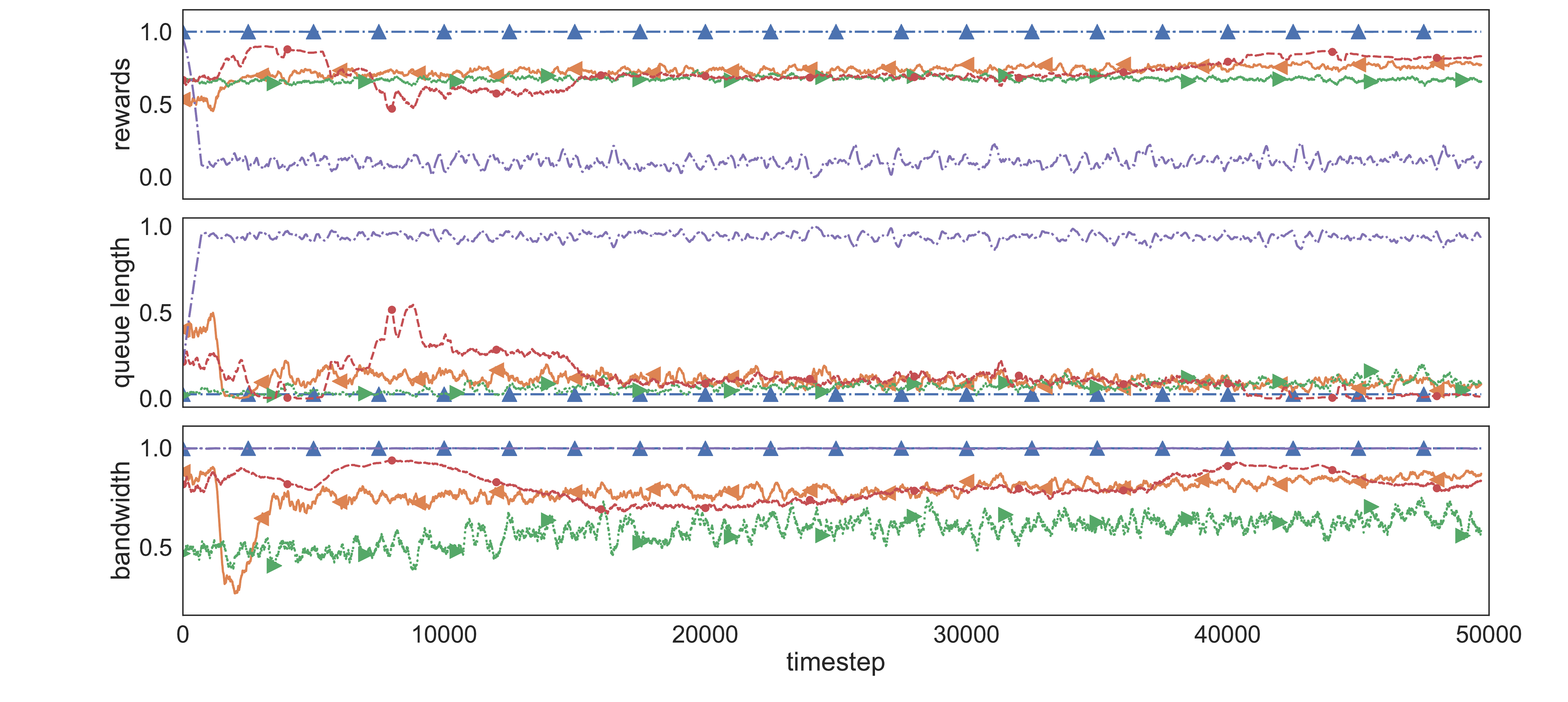}
		\caption{Algorithm performance on a TCP dumbbell topology.}
		\label{fig:dumbbell_tcp}
	\end{subfigure}
	%	\caption{Algorithm performance on dumbbell topology.}
	\vspace{-15px}
\end{figure}

%%%%%%%%%%%%%%%%%%%%%%%%%%%%%%%%%%%%%%%%%%%%%%%%%%
\section{Preliminary Experiments}
%%%%%%%%%%%%%%%%%%%%%%%%%%%%%%%%%%%%%%%%%%%%%%%%%%

As preliminary analysis we used Iroko to compare the performance of three established deep 
RL algorithms: REINFORCE~\cite{Williams1992REINFORCE, sutton2000Policy}, the 
Proximal Policy Gradient (PPO)~\cite{ppo}, and Deep Deterministic Policy 
Gradient (DDPG)~\cite{ddpg}. 

\begin{wrapfigure}{r}{0.35\linewidth}
	\centering
	\includegraphics[width=\linewidth]{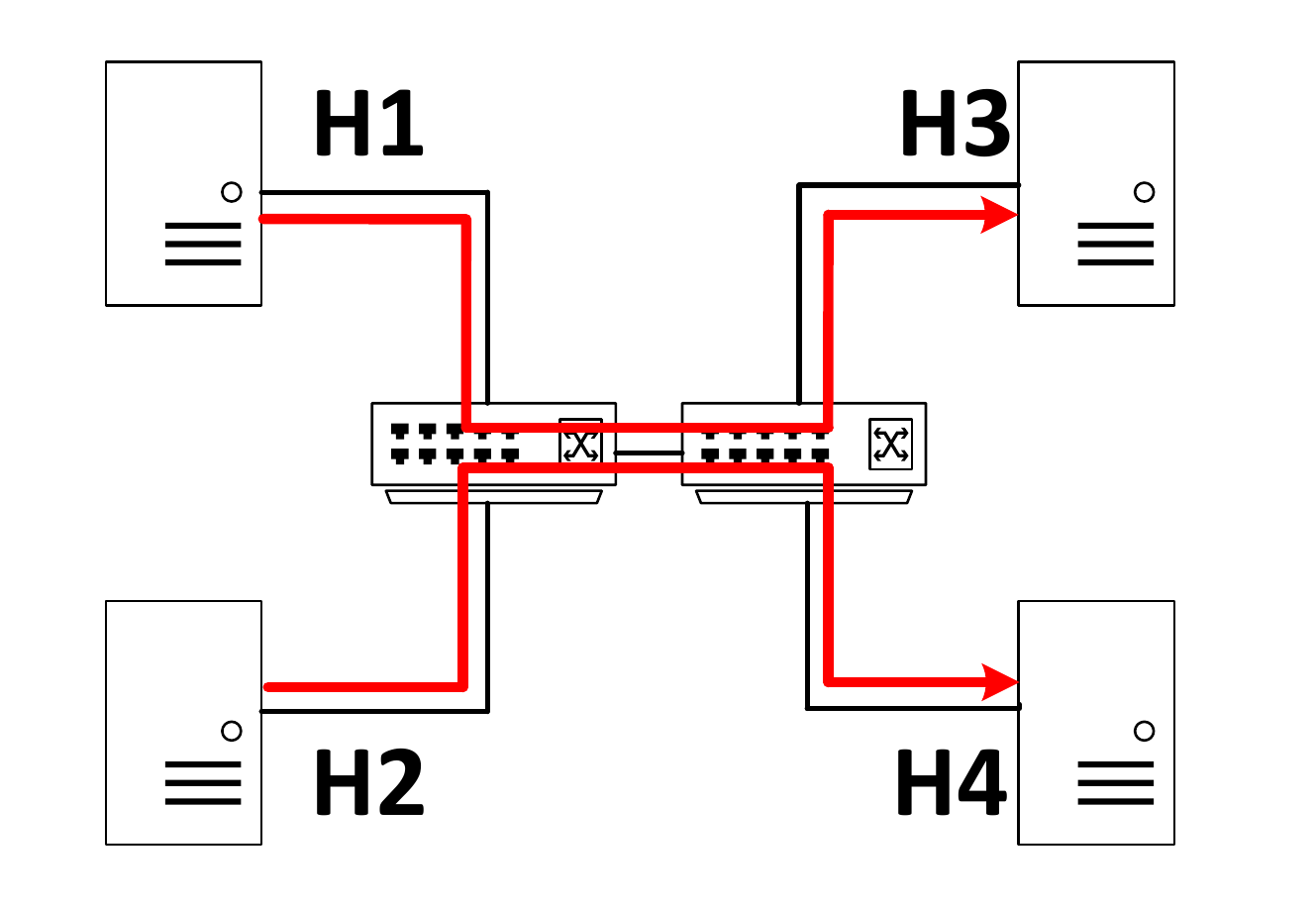}
	\caption{The dumbbell scenario.}
	\label{fig:dumbbell}
	\vspace{-10pt}
\end{wrapfigure}

We use the RLlib implementations of the three deep RL algorithms. The
library is still growing, so for REINFORCE and PPO we used the default
configurations. For DDPG, we chose the parameters to align
closely with the original work configuration, except for adding batch
normalization. We flatten the collected state into a fully connected
neural network architecture (this is an approach similar
to~\cite{auto} and~\cite{ml_resource}). Although choosing appropriate
hyper-parameters can drastically affect algorithm
performance~\cite{Henderson2018DeepRL}, we leave tuning to future
work. The full configuration details of the algorithms and hardware
specifications of our setup are listed in the Appendix.

Our first benchmark uses a dumbbell topology with four hosts connected
over two switches and a single 10 Mbit link (Figure~\ref{fig:dumbbell}). Hosts 
\texttt{H1} and \texttt{H2} are
sending constant 10 Mbit traffic to the hosts on the opposite side,
causing congestion on the central link. A trivial and fair solution to
this scenario is an allocation of 5 Mbit for each host pair.  As
comparison, we run a separate RL environment where all CC decisions
are strictly managed by TCP. We compare the RL policies against TCP
New Vegas~\cite{tcp_new_vegas} and Data Center TCP
(DCTCP)~\cite{dctcp}.

\begin{wrapfigure}{r}{.5\linewidth}
	\centering
	\includegraphics[width=\linewidth]{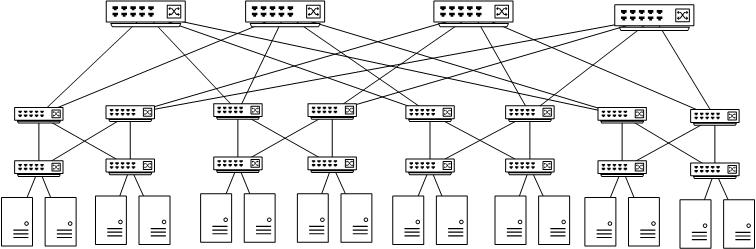}
	\caption{The fat-tree scenario.}
	\label{fig:fattree}
	\vspace{-10pt}
\end{wrapfigure}

TCP New Vegas and DCTCP are state-of-the-art congestion avoidance 
algorithms optimized for DC traffic. DCTCP requires Active Queue Management 
(AQM)~\cite{aqm} on switches, which marks packets exceeding a specific queue 
threshold with an explicit congestion notification (ECN) tag before delivering 
them. DCTCP uses this information to adjust its sending rate preemptively. 
While DCTCP is effective at avoiding congestion and queue build-up, it is 
still incapable of avoiding queues or bursts altogether~\cite{d2,d3}. In our 
experiments we treat DCTCP as the possible TCP optimum and TCP New Vegas as a 
conventional TCP baseline.

For the TCP algorithms the RL policy is ignored, but the same reward function 
is recorded. This serves as a baseline comparison, and provides empirical evidence 
on the behavior of a classical CC scheme viewed through the lens of a 
reinforcement policy.

We run each RL policy five times for 50,000 timesteps using both TCP and UDP 
as transport protocols. Each timestep is set to 0.5 seconds to give enough time to 
collect the change in queues and bandwidth in the network. A full test under 
this stepsize translates to a duration of \textasciitilde7 hours per test.

\begin{figure}[h]
	\centering
	\includegraphics[width=\linewidth]{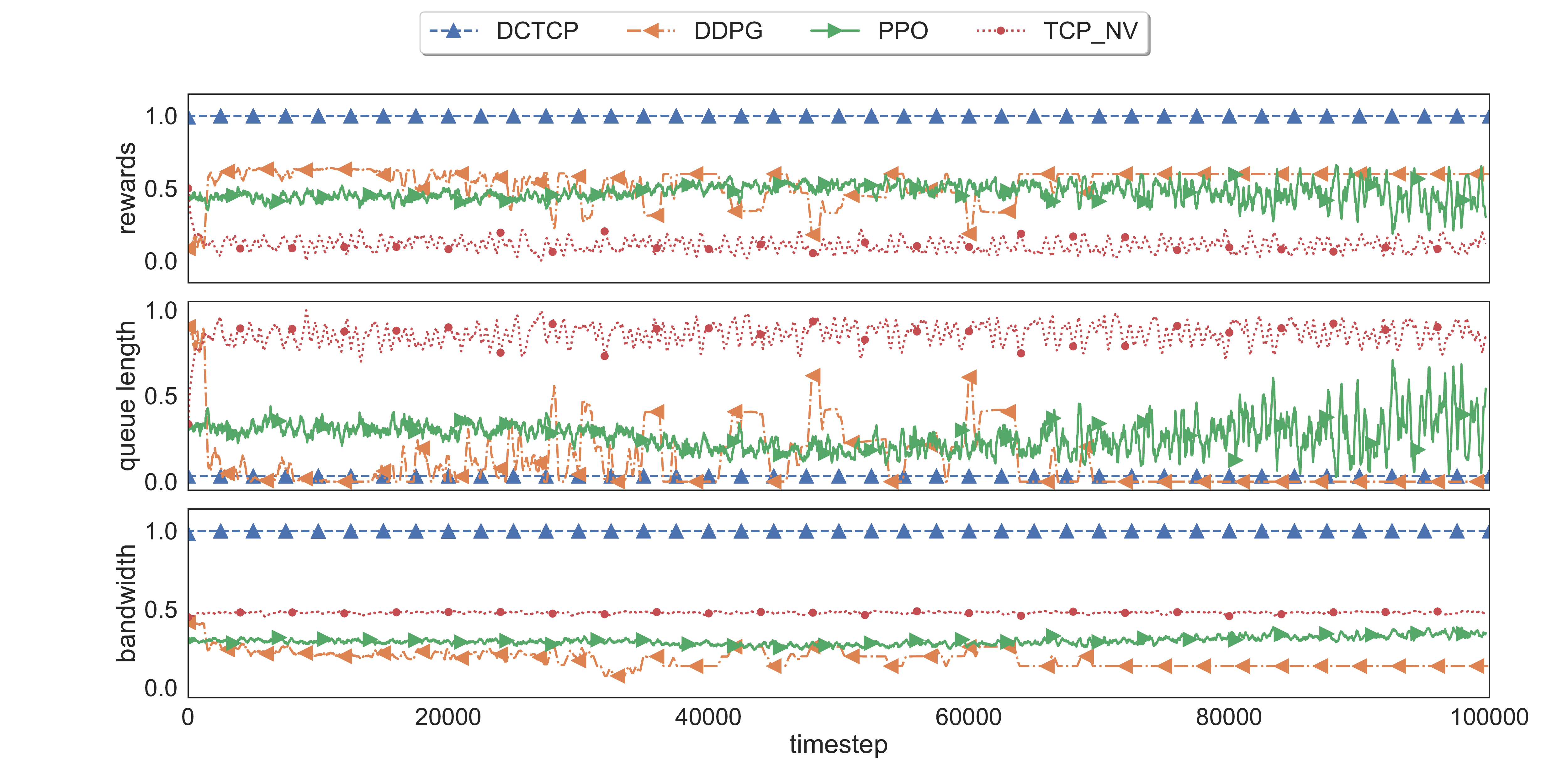}
	\caption{Algorithm performance on a fat-tree topology.}
	\label{fig:fattree_udp}
	\vspace{-15px}
\end{figure}

TCP's flow control acts as a decentralized CC agent, which is a potential 
factor in confounding the contribution of the policy learned by the RL 
algorithm. UDP does not have flow control and is not typically used in the DC 
setting, but this pushes all congestion management to the RL 
algorithm. We measure the change in average reward, the utilization of each 
host interface, and the queue build-up on the congested link.

In addition, we investigate PPO's and DDPG's performance in a more
complex DC scenario. We run the each algorithm for 100,000 timesteps
(~14 hours) on a UDP-based fat-tree 
topology~\cite{fattree} (Figure~\ref{fig:fattree} with 16
hosts and 20 switches. This results in a $80 \times 16$ state
matrix and an action vector of length 16.

\subsection{Results}

Figures~\ref{fig:dumbbell_udp} and~\ref{fig:dumbbell_tcp} plot the
dumbbell topology results for the UDP and TCP settings. We see that
DDPG achieves the highest reward and continues to improve. All
algorithms beat TCP New Vegas in reward, while minimizing the queue
buildup on the congested link. This implies that the algorithms
quickly learn a positive allocation.  Interestingly, REINFORCE
performs much better in combination with TCP. Policies such as PPO or
REINFORCE are estimated to work better in stochastic
environments~\cite{Henderson2018DeepRL}. DC environments, and TCP in
particular, exhibit stochastic characteristics (e.g., wildly varying
throughput or unstable flow behavior), which may explain the good
performance of REINFORCE.

DDPG emerges as the best choice of the three implementations. This is likely 
due to the deterministic nature of our traffic. With a more
stochastic pattern we expect to see a shift in performance in favor of PPO. 
We leave this as future work. % Trials in such a scenario are left as future work.

Figures~\ref{fig:fattree_udp} shows the fat-tree measurements. DDPG has the 
best performance, but converges to a very low bandwidth setting. PPO is very 
volatile but continuously improves in bandwidth. We believe a 
higher step count, multiple runs, and configuration tuning are required to 
produce conclusive evidence on the algorithms' performance.

Overall, however, DCTCP remains unbeaten. This is expected as DCTCP is a highly 
optimized algorithm with continuous kernel support. Our reinforcement learning 
algorithms only use a basic configurations and perform actions on a coarse 0.5 
second scale. In addition to reducing the action granularity, we are also 
investigating solutions that allow for more complex actions (e.g., providing a 
series of actions for the next n-seconds).

%A stochastic policy such as PPO has the potential to perform better than DDPG in highly fluctuating environments (such as data centers)~\cite{Henderson2018DeepRL}.

%\input{challenges}
\section{Concluding Remarks}

Deploying reinforcement learning in the DC remains challenging. The
tolerance for error is low and decisions have to be made on a
millisecond scale. Compared to a TCP algorithm on a local host, a DC
agent has to cope with significant delay in its actions. The chaotic and
opaque nature of DC networks makes appropriately crediting
actions nearly impossible.
Rewards, actions, and state can be mix-and-matched arbitrarily. There is no 
indication or theoretical insight if a particular combination will be 
successful. The fact that traffic has to be evaluated in real-time leads to 
slow prototyping and agent learning curve. Optimizing a network of a mere 16 
hosts is already a substantial task, since each node is an independent actor 
with unpredictable behavior.

Nonetheless, our initial results are encouraging. In the
dumbbell tests, the agents can quickly learn a fair distribution policy, 
despite the volatility of the network traffic. DDPG and PPO even exceed the TCP 
New Vegas baseline and demonstrate steady improvement.

We plan to continue work on our benchmarking tool and focus on improving 
the emulator performance for fat-tree scenarios. This includes hyper-parameter 
tuning and deployment automation using the Ray framework.
We are looking into using meta-information such as job deployments,
bandwidth requests by nodes, or traffic traces as additional state
information. We also plan to extend the range of reward models,
topologies, traffic patterns, and algorithms to truly evaluate the
performance of reinforcement learning policies. Iroko is an
open-source project available at \url{https://github.com/dcgym/iroko}.

\subsubsection*{Acknowledgments}
We like to thank Mark Schmidt for his feedback on this paper.
This work is supported by Huawei and the Institute for Computing,
Information and Cognitive Systems (ICICS) at UBC.
\bibliographystyle{acm}
%\bibliography{iroko}
\bibliography{bibliography}
\newpage
\section{Appendix}

All experiments were run using a Linux 4.15.0-34 kernel on a single 8 core 
(2xIntel Xeon E5-2407) machine with 32GB of RAM.  
We used Ray version 0.5.3. Network emulation is performed using the Linux 
NetEm~\cite{netem} package. All hosts are connected over instances of the Open 
vSwitch~\cite{ovs}.
The sending rate of hosts is adjusted via the following Linux 
traffic control command:\\
\texttt{tc qdisc change dev [iface] root fq maxrate [bw]mbit}\\
The monitors collect network statistics using the Linux tools 
\texttt{ifstat}, \texttt{tc qdisc show}, or \texttt{tcpdump}. Traffic is 
generated using the Goben traffic generator written in Go.  Goben is 
open-source and located at \url{https://github.com/udhos/goben}.
Each algorithm utilize a neural network model with
two hidden layers both with 256 neurons, and \texttt{tanh} for activation. The DDPG algorithm also uses this model with additional parameters for the actor and critic neural networks as specified in the table.
Hyperparameter names are written to closely follow the current variable named 
used in RLlib:

\begin{table}[h]
\centering
\begin{tabular}{|l|l|}
 \hline
Hyperparameter                                                          & 
Value                        \\  \hline
\multicolumn{2}{|c|}{DDPG} \\
\hline
$\theta$                                                                & $0.15$                       \\  
$\sigma$                                                                & $0.2$                        \\  
Noise scaling                                                                & $1.0$                        \\ 
Target network update frequency                                         & Every update                  \\  
$\tau$                                                                  & $10^{-3}$ \\
Use Prioritized Replay Buffer~\cite{Schaul2015Prioritized}              & 
False                          \\  
%$\alpha$                                                                & 0.6                          \\  
%$\beta$                                                                 & 0.4                          \\  
%temporal difference $\epsilon$ & $10^{-6}$ \\  
Actor hidden layer sizes													& 400, 300 \\
Actor activation function 												& ReLU \\
Critic hidden layers sizes													& 400, 300 \\ 
Critic activation function 												& ReLU \\
Optimizer                                                               & Adam \cite{Kingma2014Adam}   \\  
Actor learning rate                                                     & $10^{-4}$                    \\  
Critic learning rate                                                    & $10^{-3}$                    \\  
Weight decay coefficient                                                & $10^{-2}$                   \\ 
Critic loss function                                                    &  
Square loss              \\
 \hline
\multicolumn{2}{|c|}{PPO} \\
\hline
Use GAE~\cite{schulman2015High}    & True      \\  
GAE Lamda                  & 1.0       \\  
KL coefficient                  & 0.2       \\  
Train batch size           & 4000   \\  
Mini batch size            & 128       \\  
Num SGD Iterations & 30 \\
Optimizer & Adam~\cite{Kingma2014Adam}  \\  
Learning rate              & $5*10^{-5}$ \\  
Value function coefficient & 1.0       \\  
Entropy coefficient        & 0.0       \\  
Clip parameter             & 0.3       \\  
Target Value for KL        & 0.01     \\
\hline
\multicolumn{2}{|c|}{REINFORCE} \\
\hline
Learning rate               & $10^{-4}$  \\ 
Optimizer & Adam~\cite{Kingma2014Adam}  \\  \hline
\end{tabular}
\caption{Configurations for all algorithms.}
\end{table}

\end{document}